\documentclass[reprint,amsmath,amssymb,aps,prb,amssymb,superscriptaddress,nofootinbib]{revtex4-1}
\usepackage[utf8]{inputenc}

\usepackage{color}
\usepackage{braket}
\usepackage{mathtools}
\usepackage{bm}
\usepackage{amsfonts}
\usepackage{dcolumn}
\usepackage{natbib}
\usepackage{bbold}
\usepackage{soul}
\usepackage[dvipsnames]{xcolor}
\usepackage{amsmath, amssymb}
\usepackage[breaklinks=true,colorlinks,citecolor=blue,linkcolor=blue,urlcolor=blue]{hyperref}
\usepackage{graphicx}
\usepackage{float}
\usepackage{appendix}

\makeatletter
\renewcommand{\selectlanguage}[1]{}       
\makeatother

\makeatletter
\def\blfootnote{\xdef\@thefnmark{}\@footnotetext}
\makeatother

\begin{document}

\title{Atomistic origin of low thermal conductivity in quaternary chalcogenides \\Cu(Cd, Zn)$_2$InTe$_4$} 

\author{Nirmalya Jana} 
\email{nirmalyaj20@iitk.ac.in}
\affiliation{Department of Physics, Indian Institute of Technology Kanpur, Kanpur-208016, India}

\author{Amit Agarwal} 
\email{amitag@iitk.ac.in}
\affiliation{Department of Physics, Indian Institute of Technology Kanpur, Kanpur-208016, India}

\author{Koushik Pal$^\perp$} 
\email{koushik@iitk.ac.in}
\affiliation{Department of Physics, Indian Institute of Technology Kanpur, Kanpur-208016, India} 
\blfootnote{$\perp$ Corresponding author}


\begin{abstract}
Crystalline semiconductors with intrinsically low lattice thermal conductivity ($\mathcal{K}$) are vital for device applications such as barrier coatings and thermoelectrics. Quaternary chalcogenide semiconductors such as CuCd$_2$InTe$_4$ and CuZn$_2$InTe$_4$ are experimentally shown to exhibit low $\mathcal{K}$, yet its microscopic origin remains poorly understood. Here, we analyse their thermal transport mechanisms using a unified first-principles framework that captures both the Peierls (particle-like propagation, $\mathcal{K}_P$) and coherence (wave-like tunneling, $\mathcal{K}_C$) mechanisms of phonon transport. We show that extended antibonding states below the Fermi level lead to enhanced phonon anharmonicity and strong scattering of heat-carrying phonon modes, suppressing $\mathcal{K}$ in these chalcogenides. We show that $\mathcal{K}_P$ dominates the total thermal conductivity, while $\mathcal{K}_C$ remains negligible even under strong anharmonicity of the phonon modes. The heavier Cd ions in CuCd$_2$InTe$_4$ induce greater acoustic-optical phonon overlap and scattering compared to CuZn$_2$InTe$_4$, further lowering  thermal conductivity of the former. Additionally, grain boundary scattering in realistic samples contributes to further suppression of thermal transport. Our findings establish the atomistic origins of low $\mathcal{K}$ in quaternary chalcogenides and offer guiding principles for designing low-thermal-conductivity semiconductors.
\end{abstract}

\maketitle

\section{Introduction} 

Semiconducting quaternary chalcogenides such as CuCd$_2$InTe$_4$ and CuZn$_2$InTe$_4$ are important energy materials with potential applications in photovoltaic technology \cite{chen_2009, chen_2010, ghosh_2016}. Owing to their earth-abundant constituents, these materials have been widely investigated to understand the microscopic mechanisms underlying their transport processes and to enhance their photo-conversion efficiency (PCE). While charge carrier transport is directly linked to the PCE in photovoltaics, phonon-mediated heat transport, i.e., lattice thermal conductivity ($\mathcal{K}$), can also significantly influence device performance. For instance, it has been shown that very low $\mathcal{K}$ in semiconductors can give rise to the hot-phonon bottleneck effect, which prolongs the lifetime of photo-generated carriers and thereby enhances the PCE \cite{yang_2016, yang_2017, ross_1982, konig_2010, gavin_2009}. Crystalline semiconductors with intrinsically low $\mathcal{K}$ are also essential for thermal management in applications such as thermoelectrics \cite{ghosh_2022, tan_2016, chen_2021, yan_2022}, thermal barrier coatings \cite{nitin_2002}, and refractories. Therefore, understanding the microscopic origin of low $\mathcal{K}$ is not only of fundamental interest but also crucial for designing new materials with targeted technological applications.

Beyond these specific compounds, many quaternary chalcogenides possessing diverse structures, chemistry and compositions have been experimentally shown to exhibit intrinsically low lattice thermal conductivities, for example Ba$_3$Cu$_2$Sn$_3$Se$_{10}$ \cite{Ojo2022_Ba3Cu2Sn3Se10}, Cu$_2$ZnSnS$_4$ and related kesterites \cite{Sharma2019_CZTS_lowK, Mukherjee2023_CZTS_SE_lowK}, BaCu$_2$SnQ$_4$ (Q = S, Se) \cite{BaCu2SnQ4_2020}, and  In$_4$Pb$_{5.5}$Sb$_5$S$_{19}$ \cite{Li2020_In4Pb5_5Sb5S19}. Their wide composition space  offers systematic control over mass, bonding, and vibrational spectra, making them an excellent platform to design and understand low-\(\mathcal{K}\) semiconductors. Recent studies on CuCd$_2$InTe$_4$, CuZn$_2$InTe$_4$, and related chalcogenides confirm this potential and motivate a microscopic investigation of the mechanisms underlying their suppressed thermal transport~\cite{pal2021accelerated, hobbis_2018,ojo_2021}.
  
At the microscopic level, several distinct and often coexisting mechanisms have been identified in different low-$\mathcal{K}$ crystals: (i) weak or antibonding chemical interactions that reduce interatomic force constants and elastic moduli, thereby softening acoustic branches and enhancing anharmonicity~\cite{gholami_2024,das_2023}; (ii) low-lying, nearly dispersionless optical phonon branches that overlap with acoustic branches and create  resonant three-phonon channels, i.e., a large three-phonon phase space~\cite{Jana2017, Feng2017}; (iii) in some tellurides, strong quartic anharmonicity leads to significant four-phonon scattering processes that further suppress $\mathcal{K}$~\cite{feng2016,Feng2017}, though their importance is highly material dependent and computationally demanding to evaluate. Which mechanism dominates is highly material-specific, so a unified  first-principles analysis is required to identify the operative microscopic cause in any given compound \cite{simoncelli2019unified,simoncelli2022wigner,xiao2020tl3vse4,pal2021microscopic}.

 Recent experiments have uncovered ultralow thermal conductivity in the CuCd$_2$InTe$_4$ family of quaternary chalcogenide semiconductors \cite{hobbis_2018}, and ongoing efforts aim to reduce it further \cite{ojo_2021}. 
 Therefore, it is crucial to understand how factors such as crystal structure, chemical bonding, and atomic vibrational properties contribute to such low thermal conductivity. Such insights can be used to design and engineer materials with intrinsically low thermal conductivity. Theoretical investigations of $\mathcal{K}$ have traditionally focused on the particle-like propagation of phonon quasiparticles \cite{peierls_1929, srivastava_2019, ziman_2001, broido_2007, cepellotti_2016, chaput_2013, omini_1995}. However, recent theoretical and computational developments have revealed that wave-like tunneling of phonons can also significantly contribute to $\mathcal{K}$ \cite{simoncelli_2019, shenogin_2009}, especially in low-$\mathcal{K}$ crystalline compounds such as CH$_3$NH$_3$PbI$_3$ \cite{yang_2022, yang_2023}, Tl$_3$VSe$_4$ \cite{jain_2020}, TlInTe$_2$ \cite{pal2021microscopic}, Cs$_3$Bi$_2$Br$_9$ \cite{li_2024}, and Cs$_2$NaInCl$_6$ \cite{wang_2024}. Insights from these studies have led to improved materials design principles targeting ultralow $\mathcal{K}$. Nonetheless, despite their prominence, the microscopic origin of low $\mathcal{K}$ in CuCd$_2$InTe$_4$ and CuZn$_2$InTe$_4$, particularly the role of the coherent term ($\mathcal{K}_C$) associated with wave-like phonon tunnelling, remains poorly understood.\\
 Here, we perform a detailed analysis of electronic structures, phonon dispersions, and thermal conductivity of two quaternary chalcogenides, CuZn$_2$InTe$_4$ and CuCd$_2$InTe$_4$, to understand the origin of their low lattice thermal conductivity. Our calculations reveal that filled antibonding valence states just below the Fermi level enhance phonon anharmonicity, particularly in the low-energy modes, leading to strong phonon-phonon scattering. We confirm this by calculating the anharmonic three-phonon scattering rates and solving the Peierls-Boltzmann transport equation iteratively to determine $\mathcal{K}$. We decompose the total thermal conductivity into contributions from particle-like phonon propagation ($\mathcal{K}_P$) and wave-like tunnelling ($\mathcal{K}_C$), and find that $\mathcal{K}_P$ dominates over $\mathcal{K}_C$ in both materials. However, the combined effect of these mechanisms still underestimates the experimentally observed suppression of $\mathcal{K}$. To reconcile this discrepancy, we additionally account for phonon scattering at grain boundaries, which further reduces thermal transport and aligns our results with experimental observations. Our combined first-principles and quantum transport framework provides a microscopic understanding of lattice heat conduction and offers design principles for engineering semiconductors with intrinsically low thermal conductivity.

\section{Methods}
We performed all first-principles density functional theory (DFT) calculations using Vienna ab initio simulation package (VASP)~\cite{kresse1996efficient, kresse1999ultrasoft}. We used projector-augmented wave (PAW) potentials~\cite{paw94, paw99}, and the generalised gradient approximation (GGA) for the exchange-correlation functional as implemented by the revised Perdew-Burke-Ernzerhof functional for solids (PBEsol)~\cite{perdew_2008}. 
Valence electronic configuration of the atoms used in the pseudopotentials are 3d$^{10}$4s$^{1}$ for Cu, 4d$^{10}$5s$^{2}$ for Cd, 5s$^{2}$5p$^{1}$ for In, 5s$^{2}$5p$^{4}$ for Te, and 3d$^{10}$4s$^{2}$ for Zn.

The quaternary chalcogenides CuCd$_2$InTe$_4$ and CuZn$_2$InTe$_4$ crystallise in the modified Zinc blende structure \cite{nolas_2016}. We constructed three possible configurations of each compound and fully optimised the structures until the Hellman-Feynman forces on each atom were less than  10$^{-4}$ eV/\AA. The 
lowest-energy structure for each quaternary chalcogenide was used for subsequent calculations. We chose 400 eV as the kinetic energy cutoff for the plane wave basis and $11\times11\times 11$ $k$-mesh for the self-consistent field (scf) calculations. 

To determine bulk and shear moduli, we calculated the elastic constants using the finite difference method with symmetry constraints as implemented in VASP. We analysed  chemical bonding in the studied compounds using crystal orbital Hamiltonian populations (COHP)\cite{dronskowski_1993} analysis, which is localised pairwise energy contributions of crystal orbitals to the one-particle Bloch bands. We represent the localised orbitals on each atom by Slater-type orbitals\cite{nelson_2020}, chosen to match the valence orbitals used in our DFT calculations. The projected COHP between two localised orbitals $\ket{\phi_{\mu}}$ and $\ket{\phi_{\nu}}$ is given by, COHP$_{\mu\nu}(E, \textbf{k}) = \sum_i \textbf{Re}[\braket{\phi_{\mu}|\psi_i(\textbf{k})}\braket{\psi_i(\textbf{k})|\phi_{\nu}}$ $\sum_j \braket{\phi_{\nu}|\psi_j(\textbf{k})}\epsilon_j(\textbf{k})\braket{\psi_j(\textbf{k})|\phi_{\mu}}]\times \delta(\epsilon_i(\textbf{k})-E)$ \cite{deringer_2011}. Here, $E$ and $\textbf{k}$ denote the energy and crystal momentum being probed, $\ket{\psi_i(\textbf{k})}$ is the Bloch state of the $i^{th}$ band with energy $\epsilon_i(\textbf{k})$. We obtained COHP$(E)$ for a pair of atoms by summing COHP$_{\mu\nu}(E, \textbf{k})$ over the valence orbitals ($\ket{\phi_{\mu}}$) localised on the atomic sites and integrating over $\textbf{k}$-points in the Brillouin zone, as implemented in the LOBSTER program \cite{dronskowski_1993, maintz_2016, nelson_2020}.

\begin{figure*}[]
    \centering
    \includegraphics[width=0.95\linewidth]{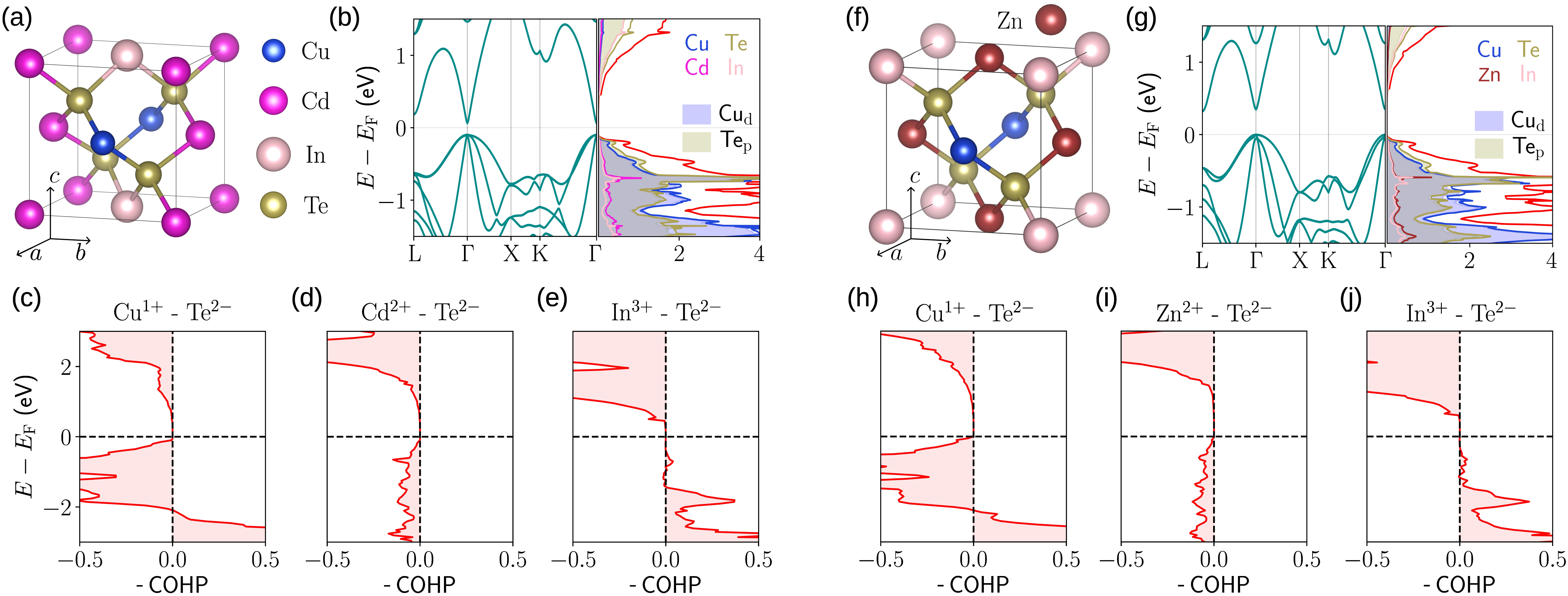}
    \caption{Zinc-blende like crystal structure of (a) CuCd$_2$InTe$_4$, (f) CuZn$_2$InTe$_4$. (b, g) Electronic  structures (left) and density of states (DOS) (right) show the semiconducting gap for both materials. The crystal orbital Hamilton population (COHP) in figures (c, d, e, h, i, j) represent the bonding (+ve) and antibonding (-ve) nature between any pair of nearest neighbour cation and anion. Large DOS of Cu and Te ions with a significant contribution from Cd/Zn and their antibonding states below the Fermi energy for both the compounds influence the thermal transport.} 
    \label{fig1}
\end{figure*}
We computed the second-order interatomic force constants (IFCs) using the finite difference method within phono3py \cite{phono3py, phonopy-phono3py-JPCM} using a 2$\times$2$\times$2 supercell. Third-order IFCs were evaluated by including up to the sixth nearest neighbours $\sim 5.4$ {\AA}) interaction using the same 2$\times$2$\times$2 supercell. We verified the convergence of third-order interatomic force constants with respect to the cutoff radius by comparing results obtained using cutoff distances of 4.5 Å (4th nearest neighbors) and 5.4 Å (6th nearest neighbors). The lattice thermal conductivities computed with both cutoffs show excellent agreement within numerical uncertainty (see Appendix A, Fig.~\ref{fig1_a}b), confirming that the chosen cutoff of 5.4 Å provides reliable and converged results. This is consistent with previous reports for three-dimensional bulk materials \cite{Carrete2015}. 
We calculated the forces on the atoms using a $\Gamma$-centred 2$\times$2$\times$2 $k$-mesh. We calculated the lattice thermal conductivity by solving the linearised Boltzmann transport equation (LBTE) for phonons \cite{phono3py_lbte} and the Wigner transport equation \cite{phono3py_wigner, simoncelli_2019}, as implemented in Phono3py \cite{phono3py_lbte}, to obtain $\mathcal{K}_P$ and $\mathcal{K}_C$, respectively. For the calculations of phonon linewidth, we employed a $q-$mesh of 11$\times$11$\times$11 along with a Gaussian smearing width $\sigma \sim$ 0.05 eV to perform integration over the Brillouin zone. We explicitly account for grain boundary effects by employing Matthiessen’s rule for phonon relaxation times, as implemented in the Phono3py \cite{phono3py_lbte}. In Matthiessen’s rule for phonon scattering rates, the total phonon scattering rate, expressed in terms of the inverse relaxation time $\tau_\lambda^{-1}$, is given by
\begin{equation}
    \tau_\lambda^{-1} = \frac{1}{\tau_\lambda^{3ph}} + \frac{1}{\tau_\lambda^{gb}}~.
\end{equation}
Where, ${1}/{\tau_\lambda^{gb}} = {|v_{g, \lambda}|}/{L}$ is the phonon-grain boundary scattering rate, $v_{g, \lambda}$ is the phonon group velocity for mode $\lambda$, and $L$ is the grain size.

To quantify the degree of anharmonicity of the phonon modes, we calculated the mode Gr\"uneisen parameter ($\gamma(\textbf{q}, \nu)$) utilising the derivative of the dynamical matrix with respect to the unit cell volume \cite{zhu_2024}. $\gamma(\textbf{q}, \nu)$ of a phonon mode with wave vector $\textbf{q}$ and band index $\nu$ is given by $\gamma(\textbf{q}, \nu) = -[V/(2\omega^2(\textbf{q}, \nu))][\braket{e(\textbf{q}, \nu)|\partial D(\textbf{q})/\partial V|e(\textbf{q}, \nu)}]$. Here, $V$ and $\omega(\textbf{q}, \nu)$ are the volume of the unit cell and frequency of the phonon mode, $D(\textbf{q})$ is the dynamical matrix, and the $e(\textbf{q}, \nu)$ is the eigenvector.

Lattice thermal conductivity was calculated by including both the particle-like phonon propagation (Peierls’ picture) and wave-like tunnelling of phonons (coherences) between phonon branches\cite{simoncelli_2019, zheng_2024}. The heat flux along $\alpha$-axis is $J^{\alpha} \sim \mathcal{K}^{\alpha \beta} (\nabla T)^{\beta}$, where $(\nabla T)^{\beta}$ is the temperature gradient along $\beta$-axis. The total thermal conductivity is given by, 
\begin{equation}
    \mathcal{K}^{\alpha \beta} = \mathcal{K}^{\alpha \beta}_{P} + \mathcal{K}^{\alpha \beta}_C~.
\end{equation}
The Peierls’s contribution $\mathcal{K}^{\alpha \beta}_{P}$ describes semiclassical phonon propagation, while $\mathcal{K}^{\alpha \beta}_C$ accounts for off-diagonal coherence terms captured by the Wigner distribution function~\cite{frensley_1990, kane_2012,simoncelli_2019}. 
The coherence contribution to the lattice thermal conductivity is given by, 
\begin{eqnarray}
    \mathcal{K}_C^{\alpha \beta} = \frac{\hbar^2}{\mathrm{V_0} N_c k_{\rm B} T^2}\sum_{\textbf{q}}\sum_{s\neq s'}\frac{\omega(\textbf{q})_s+\omega(\textbf{q})_{s'}}{2}V_{s, s'}^{\alpha}(\textbf{q})V_{s', s}^{\beta}(\textbf{q})\nonumber \\
    \times \frac{\omega(\textbf{q})_s \Bar{N}(\textbf{q})_s[\Bar{N}(\textbf{q})_s + 1]+\omega(\textbf{q})_{s'} \Bar{N}(\textbf{q})_{s'}[\Bar{N}(\textbf{q})_{s'} + 1]}{4[\omega(\textbf{q})_s - \omega(\textbf{q})_{s'}]^2 + [\Gamma(\textbf{q})_s + \Gamma(\textbf{q})_{s'}]^2} \nonumber \\
    \times [\Gamma(\textbf{q})_s + \Gamma(\textbf{q})_{s'}]~.
    \label{eqn2}
\end{eqnarray}
Here, the summation runs over wavevector $\textbf{q}$ and phonon modes $s, s'$. $\omega$ is the phonon frequency, $V_{s, s'}^{\alpha}(\textbf{q})$ is the generalized phonon group velocity along the $\alpha$-axis. $\Bar{N}(\textbf{q})_s$ is the equilibrium Bose-Einstein distribution, $\Gamma(\textbf{q})_s$ is the phonon linewidth. $\hbar, \mathrm{V_0}, N_c, k_{\rm B}$ and $T$ are Planck's constant, unit cell volume, number of unit cells in the lattice, Boltzmann constant, and absolute temperature, respectively.

\begin{figure*}[]
    \centering
    \includegraphics[width=0.7\linewidth]{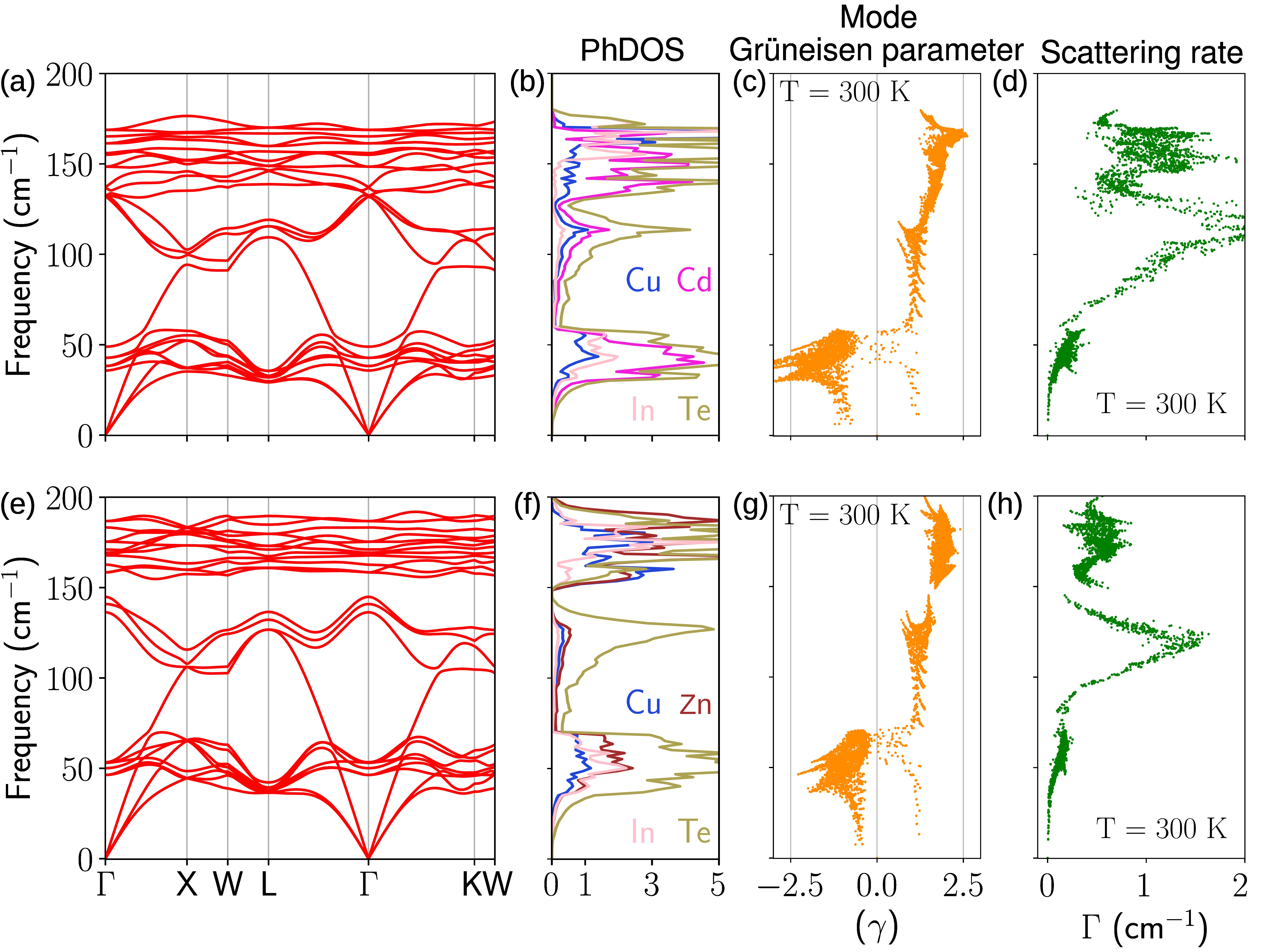}
    \caption{The upper and lower rows depict the results for CuCd$_2$InTe$_4$ and CuZn$_2$InTe$_4$, respectively. The phonon dispersions (a, e) and atom-projected phonon DOS (b, f) are shown, respectively. The mode Gr\"{u}neisen parameters ($\gamma$) (c, g) and the mode resolved phonon scattering rates (d, h) are shown for 300 K. Softer phonon modes due to heavier Cd atoms and the presence of antibonding states leads to high mode Gr\"{u}neisen parameters and large scattering rates.
    }
    \label{fig2}
\end{figure*}

\section{Results and Discussions}
\subsection{Electronic properties}
CuCd$_2$InTe$_4$ and CuZn$_2$InTe$_4$ crystallize in a modified Zinc-blende-like structure \cite{nolas_2016} as shown in Fig.~\ref{fig1}a, where the cations, Cu$^{1+}$, Zn$^{2+}$ (or, Cd$^{2+}$),  and In$^{3+}$ are bonded with four Te$^{2-}$ anions in a tetrahedral coordination environment. The calculated electronic structure of CuCd$_2$InTe$_4$ (see Fig. \ref{fig1}b) and CuZn$_2$InTe$_4$ (Fig.~\ref{fig1}g) exhibit a finite band gap of $\sim$ 0.16 eV and 0.32 eV, respectively, indicating their semiconducting nature.  The orbital-resolved density of states reveals that Cu 3d and Te 4p orbitals predominantly contribute to the valence bands below the Fermi energy ($E_{\mathrm{F}}$) in both compounds (see Fig.~\ref{fig1}b, g), with relatively smaller contributions from Cd/Zn and In atoms. These occupied valence states determine their bonding characteristics, which in turn influence the phonon modes and the thermal conductivity $\mathcal{K}$. To probe these interactions, we analysed the chemical bonding using the COHP method, which quantifies bonding, antibonding, and nonbonding contributions between atomic pairs. 

Strong hybridisation between Cu and Te orbitals pushes the occupied antibonding state close to $E_{\mathrm{F}}$ \cite{gholami_2024}. Our COHP analysis \cite{dronskowski_1993, deringer_2011} 
for different atomic pairs is shown in Fig.~\ref{fig1}(c-e) for CuCd$_2$InTe$_4$ and Fig.~\ref{fig1}(h-j) for CuZn$_2$InTe$_4$, where positive and negative values of $-$COHP correspond to bonding and antibonding orbitals, respectively. In the energy range of -2 to 0 eV, the Cu$^{1+}$ and Te$^{2-}$ exhibit strongly coupled antibonding states as shown in Figs.~\ref{fig1}c, h. In contrast, the antibonding states between Cd$^{2+}$ or Zn$^{2+}$ and Te$^{2-}$ are comparatively weaker (see Figs.~\ref{fig1}d,i).

Our analysis reveals that Cu 3d and Te 4p-orbitals hybridise strongly due to the presence of Cu at the tetrahedral environment of Te atoms, where the local inversion symmetry is broken \cite{he_2022}, giving rise to extended antibonding states \cite{das_2023}. In contrast, In$^{3+}$ and Te$^{2-}$ pair shows bonding nature (see Fig.~\ref{fig1}e, j). The extended antibonding states reduce the interatomic force constants and hence, low elastic moduli. Specifically, the bulk (B) and shear modulus (G) for CuCd$_2$InTe$_4$ (CuZn$_2$InTe$_4$) are 47 GPa (51 GPa) and 17 GPa (22 GPa), respectively. These values are lower than those in transition metal dichalcogenide WSe$_2$ (B=64 GPa and G=53 GPa)\cite{feng_2014}, a known low $\mathcal{K}$ material \cite{wang_2022}. This reduction in elastic stiffness leads to the softening of the phonon frequencies and group velocities, thereby lowering the thermal conductivity \cite{ferreira_2018, ubaid_2024, acharyya_2023}.

\subsection{Vibrational properties}
Phonon dispersions for CuCd$_2$InTe$_4$ and CuZn$_2$InTe$_4$ are shown in Figs.~\ref{fig2}a and \ref{fig2}e, respectively. Both compounds display similar low-frequency features (below 60 cm$^{-1}$), although CuCd$_2$InTe$_4$ shows softer acoustic and optical phonon modes due to the heavier mass of the Cd atom. This trend is confirmed by the group velocity-projected phonon dispersion (see Appendix A, Figs.~\ref{fig2_a}c, \ref{fig2_a}d).  The corresponding atom projected phonon density of states (PhDOS) shown in Figs.~\ref{fig2}b and \ref{fig2}f, highlights localised modes arising from Cd and Zn atoms around 50 cm$^{-1}$. This localisation is more pronounced for Cd than for Zn in CuCd$_2$InTe$_4$ and CuZn$_2$InTe$_4$, respectively. Furthermore, we find that the phonon modes are distributed mainly over three regions for both materials--CuCd$_2$InTe$_4$(0 - 60 cm$^{-1}$, 60 - 130 cm$^{-1}$, 130 - 180 cm$^{-1}$) and   CuZn$_2$InTe$_4$ (0 - 75 cm$^{-1}$,  75 - 150 cm$^{-1}$, 150 - 200 cm$^{-1}$). Acoustic modes contribute mostly at the bottom ($<$ 40 cm$^{-1}$) in the first region. 

The first region also contains low frequency and nearly dispersionless optical phonon branches (around 50 cm$^{-1}$), which are expected to give rise to enhanced phonon scattering channels and scattering rates, and thus strongly influence  $\mathcal{K}$ \cite{pal_2019}. These optical branches lead to a steep rise in PhDOS, with Te contributing most significantly in both compounds. The heavier Cd atom contributes more significantly to the PhDOS than Zn. The larger projected PhDOS of an element that participates in the antibonding states below the Fermi level leads to greater phonon anharmonicity \cite{acharyya_2023}, particularly in CuCd$_2$InTe$_4$. This is reflected in the larger mode Gr\"uneisen parameters ($\gamma(\textbf{q}, \nu)$'s) \cite{kennedy_2007} for CuCd$_2$InTe$_4$ in Fig.~\ref{fig2}c compared to CuZn$_2$InTe$_4$ in Fig.~\ref{fig2}g. The anharmonicity is more pronounced in the low-lying, nearly flat optical phonon branches, which are also expected to increase the scattering rates. 
This is further supported by explicit calculations of phonon-phonon scattering rates, shown in Fig.~\ref{fig2}d, \ref{fig2}h for  CuCd$_2$InTe$_4$ and CuZn$_2$InTe$_4$, respectively. To further quantify the available channels for phonon scattering, we have computed the three-phonon scattering phase space (P$_3$) at 300 K for both compounds. As shown in Fig.~\ref{fig2_a}a, b, CuCd$_2$InTe$_4$ exhibits a noticeably larger phase space in the low-frequency acoustic region compared to CuZn$_2$InTe$_4$. This enhanced phase space directly supports the observed higher scattering rates and lower thermal conductivity in CuCd$_2$InTe$_4$.

The second region ($\sim$ (60 - 130) cm$^{-1}$ for CuCd$_2$InTe$_4$ and $\sim$ (75 - 150) cm$^{-1}$ for CuZn$_2$InTe$_4$) features both dispersive and nearly flat optical phonon branches. The contribution of heavier Cd (see Fig.~\ref{fig2}b) in the phonon density of states is more prominent than Zn (Fig.~\ref{fig2}f), while the chalcogen atoms dominate the PhDOS in this region for both systems. The anharmonicity of the optical phonons is captured in the large $\gamma(\textbf{q}, \nu)$ value ($>$ 1), as shown in Figs.~\ref{fig2}c, \ref{fig2}g. These strongly anharmonic optical phonons have large scattering rates (see Fig.~\ref{fig2}d, \ref{fig2}h). The large contributions of Cd and Cu, which form antibonding states with Te, lead to higher scattering rates. In comparison, CuZn$_2$InTe$_4$ has lower scattering rates due to the relatively weaker contribution of Zn and Cu in that frequency range. The scattering rates increase rapidly with the increase in PhDOS, i.e., with the increase in the flatness of the optical phonon branches as more and more scattering channels become available for phonons. 

In the third region (frequency $>$ 130 cm$^{-1}$ for CuCd$_2$InTe$_4$ and frequency $>$ 150 cm$^{-1}$ for CuZn$_2$InTe$_4$), all the optical phonon branches stay nearly flat. As a consequence, these modes have very small phonon group velocities (see Appendix A, Figs.~\ref{fig2_a}c, ~\ref{fig2_a}d). Such low group velocity optical modes lead to a larger PhDOS (as shown in Figs.~\ref{fig2}b, f) with higher Te contributions and relatively smaller but significant Cd/Zn, Cu, and In contributions. The anharmonicity of the phonon modes in this region is weaker compared to the first region, as shown in Figs.~\ref{fig2}c, \ref{fig2}g. Although the scattering rates are enhanced by the larger PhDOS in this region, the very low group velocities of the phonon modes lead to a very small contribution to the particle-like component of the thermal conductivity. 

A comparative analysis of CuZn$_2$InTe$_4$ and CuCd$_2$InTe$_4$ reveals that, while both compounds exhibit similar overall thermal transport trends, the underlying vibrational properties differ in ways that substantively affect phonon scattering. In CuCd$_2$InTe$_4$, the larger contribution of Cd-derived modes at low frequencies results in a higher projected phonon DOS (see Fig.~\ref{fig2}b, f), which amplifies resonant scattering between acoustic and low-lying optical modes. This leads to higher mode Gr\"uneisen parameters and intensified intrinsic phonon-phonon interactions relative to CuZn$_2$InTe$_4$.

\subsection{Thermal transport}
\begin{figure}[]
    \centering
    \includegraphics[width=0.75\linewidth]{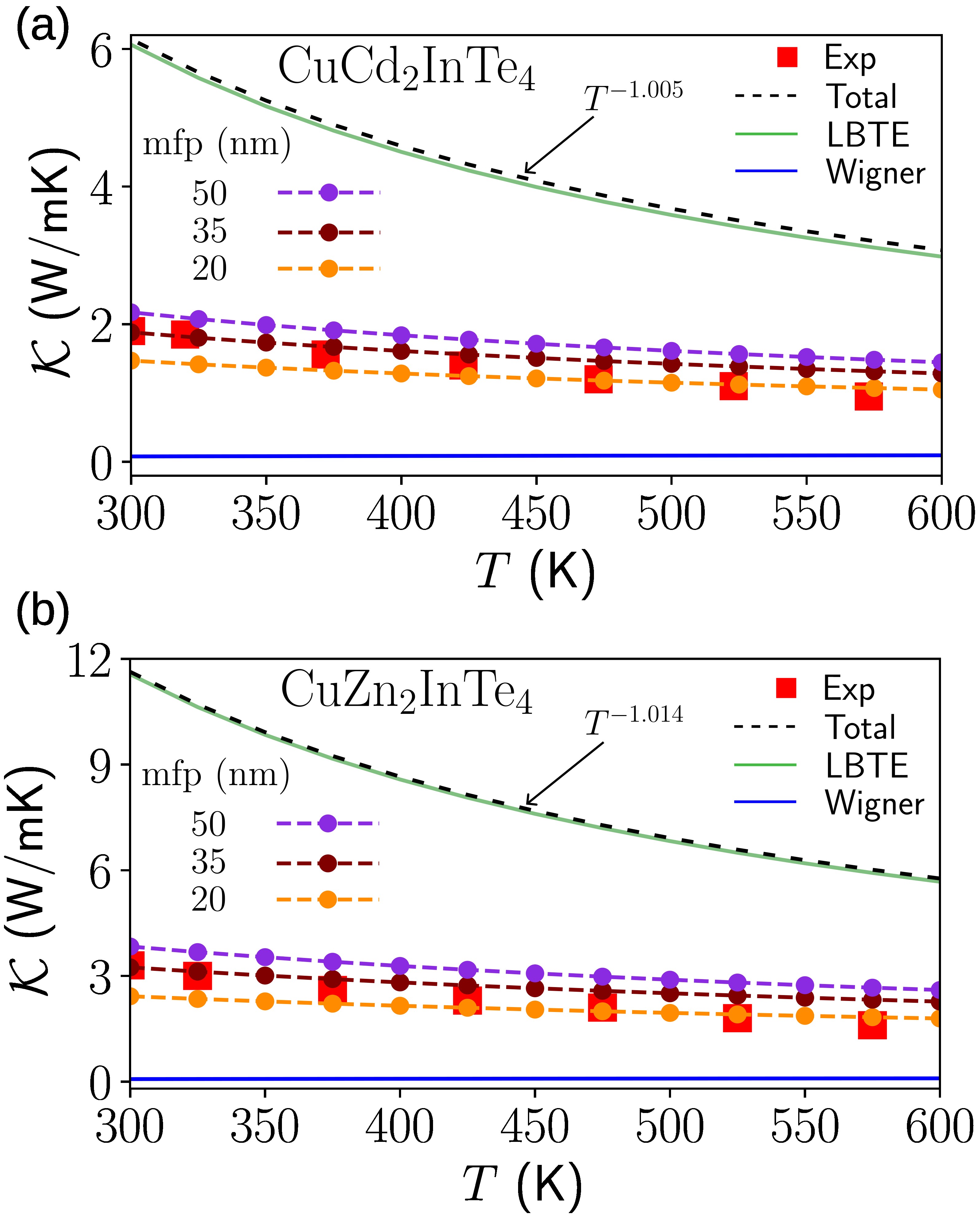}
    \caption{Calculated lattice thermal conductivity  for (a) CuCd$_2$InTe$_4$ and (b) CuZn$_2$InTe$_4$ at different temperatures. LBTE and Wigner represent the particle-like contribution ($\mathcal{K}_P$), coherences contribution ($\mathcal{K}_C$), respectively. The total thermal conductivity is the sum of both these terms, i.e., $\mathcal{K}$ = $\mathcal{K}_P$ + $\mathcal{K}_C$. The grain boundary sizes (diameters) are mentioned as mean free paths (mfp) by the numbers in nm units. CuCd$_2$InTe$_4$ shows lower thermal conductivity as predicted by our theoretical analysis, and our results also support the experimental findings for both the compounds.}
    \label{fig3}
\end{figure}

\begin{figure}[]
    \centering
    \includegraphics[width=0.65\linewidth]{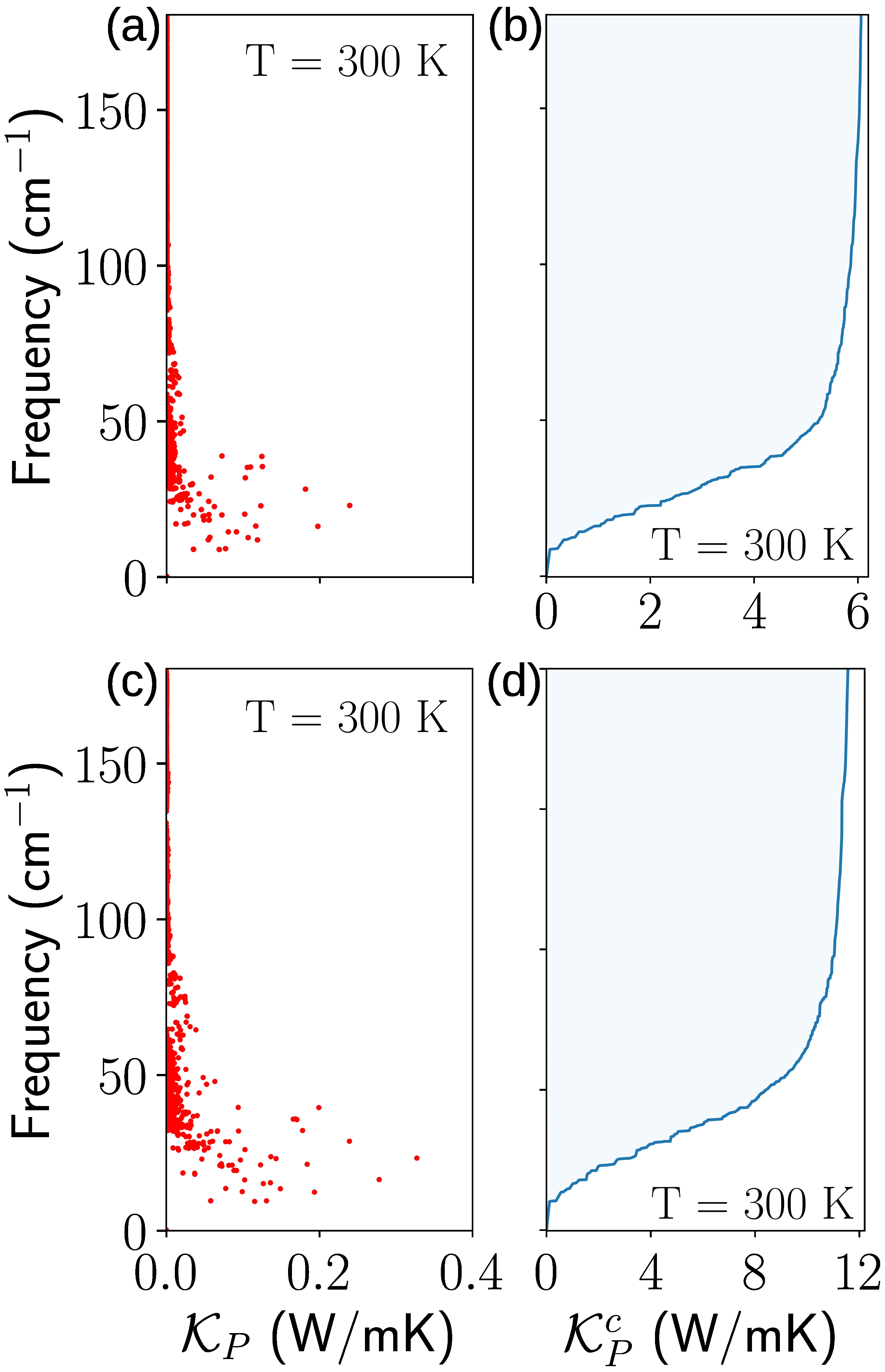}
    \caption{The upper and lower rows depict the results for CuCd$_2$InTe$_4$ and CuZn$_2$InTe$_4$, respectively. (a, c) The mode resolved particle-like contribution to thermal conductivity ($\mathcal{K}_P$) by solving LBTE. (b, d) The cumulative particle-like contribution to thermal conductivity ($\mathcal{K}_P^c$). Mode resolved contributions are dominating in the first region of phonon dispersions [CuCd$_2$InTe$_4$ (0 - 60 cm$^{-1}$) and CuZn$_2$InTe$_4$ (0 - 75 cm$^{-1}$)].}
    \label{fig1_si}
\end{figure}

Building upon the vibrational differences highlighted above, particularly the enhanced low-frequency Cd-derived modes and increased phonon scattering phase space in CuCd$_2$InTe$_4$, we now examine their impact on lattice thermal conductivity. 
We calculate the particle-like contribution (i.e., $\mathcal{K}_P$) to thermal conductivity by solving the LBTE for phonons. This contribution is shown by the green curves for CuCd$_2$InTe$_4$ and CuZn$_2$InTe$_4$ in Figs.~\ref{fig3}a and \ref{fig3}b, respectively. $\mathcal{K}_P$ of CuCd$_2$InTe$_4$ is lower compared to  CuZn$_2$InTe$_4$  throughout the temperature range. This is due to the relatively large $\gamma(\textbf{q}, \nu)$ values of the low-energy phonons (Figs.~\ref{fig2}c, \ref{fig2}g) and hence larger phonon scattering rates (Figs.~\ref{fig2}d, \ref{fig2}h). To understand the mode contributions to  $\mathcal{K}_P$ arising from different regions of phonon dispersion, we analyse $\mathcal{K}_P$ as a function of phonon frequencies and plot the cumulative values $\mathcal{K}_P^c$ in Fig.~\ref{fig1_si}. For both compounds,  phonon frequencies up to 60 cm$^{-1}$ (CuCd$_2$InTe$_4$) and 75 cm$^{-1}$ (CuZn$_2$InTe$_4$) primarily participate in the heat transport processes. As shown before, these phonon modes belong to the first region of phonon dispersion, consisting of acoustic and low-energy optical phonon branches. The antibonding valence states just below the Fermi level weaken the bonding strength. This softens the low-energy phonon frequencies and enhances the overlap of the acoustic modes with the low-lying, nearly flat optical phonon branches, resulting in large scattering rates, leading to a significant variation in $\mathcal{K}_P^c$ in the first region. Although the phonon modes in the second region of phonon dispersions exhibit high scattering rates, they contribute weakly to $\mathcal{K}_P$. This is captured by the small mode-resolved particle-like contributions and the small change of the cumulative particle-like contribution $\mathcal{K}_P^c$ (see Fig.~\ref{fig1_si}). 

The coherence contributions (i.e., $\mathcal{K}_C$) for these quaternary chalcogenides are very small (of the order of 0.1 W/mK). This is despite the strong anharmonicity of the phonon modes as revealed by their $\gamma(\textbf{q}, \nu)$ values, and shown by the blue lines in Figs.~\ref{fig3}a, \ref{fig3}b respectively. Small values of $\mathcal{K}_C$ originate from the small contributions of the off-diagonal ($s\neq s'$) generalised group velocities of phonons.
The dominant particle-like phonon conduction is captured in the $\sim T^{-1}$ behaviour of the total $\mathcal{K}$. However, the calculated lattice thermal conductivity, including both $\mathcal{K}_P$ and $\mathcal{K}_C$ contributions, is higher than the experimentally measured values \cite{hobbis_2018}. To explore the origin of this discrepancy between the experimental and calculated values of $\mathcal{K}$, we note that the presence of defects is quite common in quaternary chalcogenides \cite{makowska_2014, huang_2020}. Furthermore, experimental measurements are performed in polycrystalline samples \cite{hobbis_2018}, where the presence of grain boundaries is very common \cite{greuter_1990}.  Therefore, we consider grain-boundary scattering of phonons and the corresponding modification in the phonon linewidth. Our analysis shows that the resulting thermal conductivities in the presence of grains (with sizes between 20-50 nm) are in good agreement with the experimental values as shown in Figs.~\ref{fig3}a, \ref{fig3}b. 

It is well established that standard DFT-GGA methods typically underestimate  band gaps \cite{perdew_96}. For our systems, experimental band gaps are approximately 1 eV \cite{nolas_2016}. Despite this underestimation, our conclusions remain robust because (i) the electronic contribution to thermal conductivity in intrinsic semiconductors is typically very small and would decrease further with larger band gaps due to reduced intrinsic carrier concentrations, and (ii) our bonding and anharmonicity analyses depend mainly on the occupied valence states, which are accurately described by standard DFT functionals \cite{jones_89}. Thus, the lattice thermal conductivity and phonon properties  are not sensitive to the exact band gap value. 

We further note that electron-phonon coupling (EPC) can, in specific scenarios, influence lattice thermal conductivity by shortening phonon lifetimes. However, our systems are semiconductors with experimental band gaps of approximately 1 eV~\cite{nolas_2016}, indicating low intrinsic carrier concentrations. Recent theoretical and computational work shows that, in such systems, the contribution of EPC to thermal conductivity is negligible under intrinsic or lightly doped conditions~\cite{liao2015significant, Zhou2023, sun_2024}. Therefore, the calculated values of $\mathcal{K}$ in this study are robust with respect to the omission of EPC effects. 

Our calculations include second and third-order IFCs but omit fourth-order IFCs due to their high computational cost. This expense is especially prohibitive for our systems, which have relatively large unit cells and low crystalline symmetry, and are modeled to mimic the disorder structure of these compounds, significantly increasing the number of independent IFC components. Despite this limitation, the three-phonon scattering processes capture the dominant thermal resistance mechanisms, providing an adequate description of anharmonic thermal transport in these materials.

\section{Conclusions}

We show that the low lattice thermal conductivity in Cu(Cd,Zn)$_2$InTe$_4$ originates from strong phonon anharmonicity driven by antibonding states between Te$^{2-}$ and Cu$^{1+}$, Cd$^{2+}$, or Zn$^{2+}$. The overlap of flat, low-frequency optical branches (below 50 cm$^{-1}$) with acoustic phonons leads to enhanced three-phonon scattering and significantly suppresses thermal transport. Our analysis reveals that coherence contributions are negligible, and the particle-like propagation of acoustic and low-lying optical phonons below 75 cm$^{-1}$ primarily governs the thermal conductivity. The presence of filled antibonding states softens phonon modes, reduces group velocities, and increases the scattering phase space, all contributing to the suppression of $\mathcal{K}$. Furthermore, we show that incorporating phonon–grain boundary scattering is essential to reconcile theoretical predictions with experimental observations. Altogether, our study reveals the atomistic origins of the low thermal conductivity in these complex chalcogenides and provides guiding principles for designing next-generation low-$\mathcal{K}$ semiconductors.

Altogether, this study establishes the atomistic mechanisms that control heat transport in quaternary chalcogenides and demonstrates how chemical substitution and microstructural effects can be leveraged to engineer semiconductors with intrinsically low thermal conductivity for applications in thermoelectrics, thermal barriers, and energy devices.

\section{Data availability statement}
All data that supports the findings of this study are presented within the article.

\section{Acknowledgment}
K.P. thanks IIT Kanpur for funding support through an Initiation Grant. We acknowledge the high-performance computing facility at IIT Kanpur and the National Supercomputing Mission (NSM) for computational support.

\appendix
\renewcommand\thefigure{A\arabic{figure}}
\setcounter{figure}{0}  
\section{Supporting Information}

\begin{figure}[h]
    \centering
    \includegraphics[width=1\linewidth]{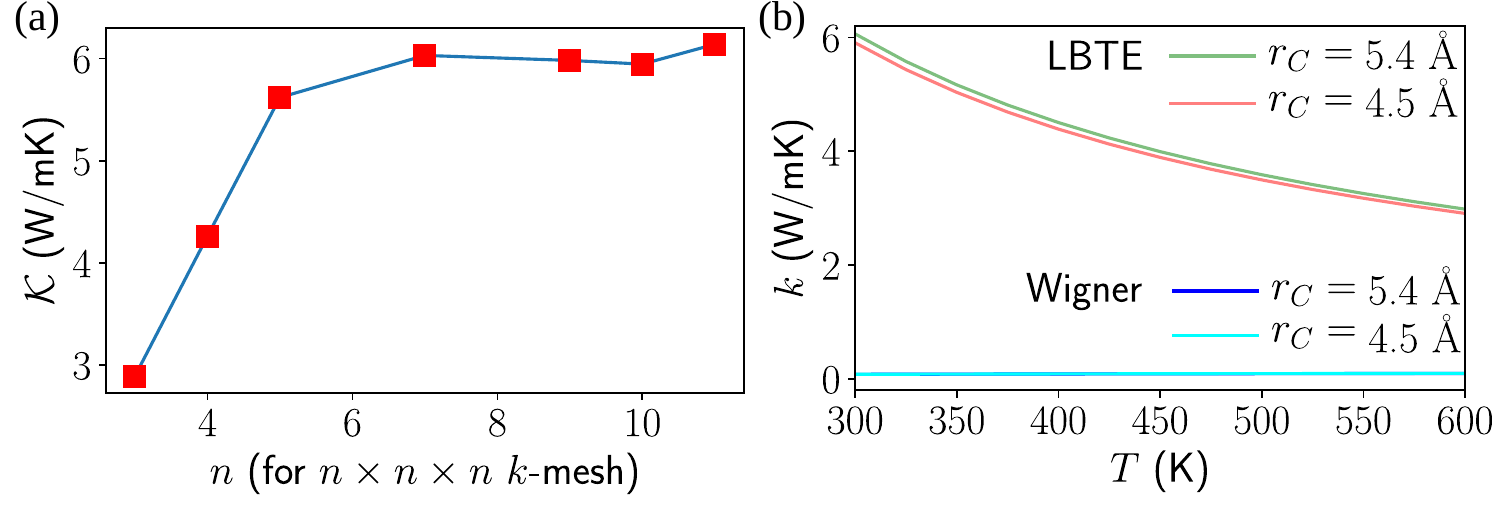}
    \caption{(a) The convergence of thermal conductivity ($\mathcal{K}$) with respect to q-mesh ($n\times n \times n$) for CuCd$_2$InTe$_4$ at a temperature of 300 K. (b) Comparison of lattice thermal conductivities (particle-like solution of linear Boltzmann transport equation (LBTE) and Wigner coherence contributions) considering two different ranges of atomic interactions. The two ranges, $r_C =$ 4.5 {\AA} and 5.4 {\AA}, contain upto four and six nearest neighbor atoms, respectively.}
    \label{fig1_a}
\end{figure}
We check the convergence of thermal conductivity with respect to $k$-mesh grid points at a temperature of 300 K. It's well converged with the $q$-points mesh grid of $11 \times 11 \times 11$ as shown in Fig.~\ref{fig1_a}a for CuCd$_2$InTe$_4$. 

\begin{figure}[h]
    \centering
    \includegraphics[width=1\linewidth]{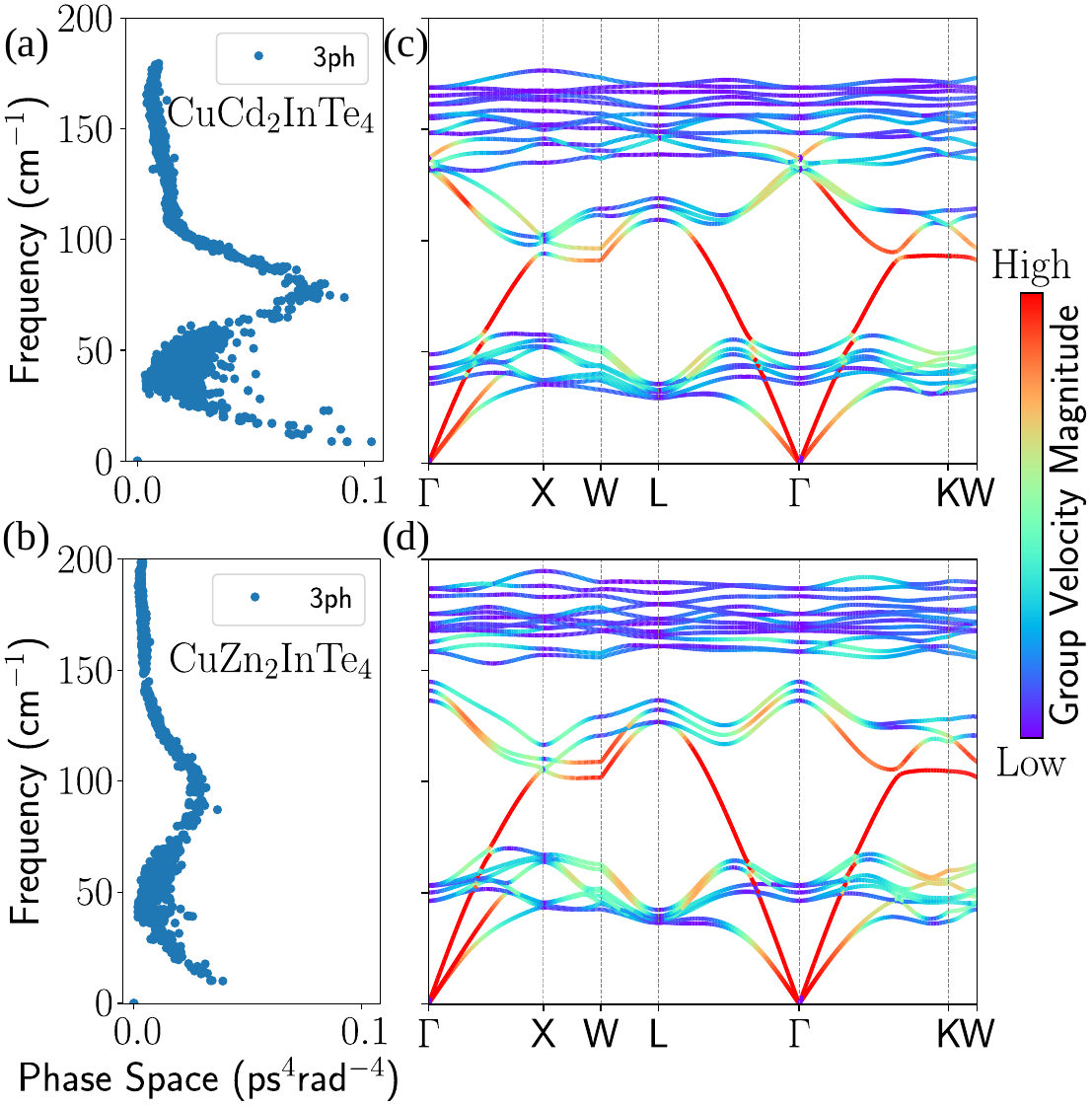}
    \caption{The three-phonon (3ph) scattering phase space (P$_3$) at 300 K for CuCd$_2$InTe$_4$ (a) and CuZn$_2$InTe$_4$ (b). Group velocity projected phonon modes along the high symmetry $k$-path for CuCd$_2$InTe$_4$ (c) and CuZn$_2$InTe$_4$ (d). The colour map highlights the magnitude of group velocity.}
    \label{fig2_a}
\end{figure}

We project the group velocity of phonon modes on the phonon band structure as shown in Fig.~\ref{fig2_a}c, d for CuCd$_2$InTe$_4$ and CuZn$_2$InTe$_4$, respectively. In the first region of phonon branches (0-60 Cm$^{-1}$ for CuCd$_2$InTe$_4$ and 0-75 Cm$^{-1}$ for CuZn$_2$InTe$_4$), the acoustic and optical modes have lower group velocities in CuCd$_2$InTe$_4$ than CuZn$_2$InTe$_4$ due to the presence of heavier Cd. We note that the identification of acoustic and optical phonon branches away from $\Gamma$ point can be challenging, especially in systems with significant mode mixing. In our phonon dispersions, we observe that the longitudinal acoustic phonon branch crosses past the low-frequency optical (40-50 cm$^{-1}$) phonon branches, while the two transverse acoustic branches remain below this region (~\ref{fig2_a}c, d). We can clearly identify these acoustic phonon branches from the phonon group velocity projected phonon dispersions, where the acoustic branches possess the highest group velocity (red) due to their linear dispersion and in-phase vibration of the atoms.
Therefore, phonon group velocity, defined as $\mathbf{v}_g = \nabla_{\mathbf{q}} \omega(\mathbf{q})$, directly relating the phonon eigenvector and its frequency, serves as a reliable proxy to distinguish acoustic-like modes from optical ones. While detailed eigenvector polarization or overlap analyses provide deeper insight, our group velocity projection provides strong evidence supporting the acoustic nature of the high-velocity mode crossing into the optical frequency range. This interpretation aligns with prior first-principles studies reporting acoustic phonon modes crossing or mixing with optical branches due to lattice complexity and mode hybridization \cite{rana2024ultralow}.

We further notice that for CuCd$_2$InTe$_4$, in the acoustic region (0–30 cm$^{-1}$), the scattering phase space (see Fig.~\ref{fig2_a}a) is large at low frequencies, starting near 0.1 ps$^4$ rad$^{-4}$ and decreasing to $\sim$0.02 ps$^4$ rad$^{-4}$ at higher frequencies within this region. The mode Gr\"uneisen parameters are predominantly negative (–1) (see Fig.~\ref{fig2}c), consistent with phonon softening and enhanced anharmonicity. Overlapping acoustic-optical region (30–60 cm$^{-1}$) shows a very high density of modes with finite phase space spanning 0.005-0.035 ps$^4$ rad$^{-4}$, with scattered points up to 0.05 ps$^4$ rad$^{-4}$. The strongly negative Gr\"uneisen parameters (–1 to –3) (see Fig.~\ref{fig2}c) indicate enhanced mode mixing and anharmonicity, increasing the scattering phase space in line with mechanisms described in Feng et al. \cite{Feng2017}. In the high acoustic branch region (60–90 cm$^{-1}$), a high group velocity acoustic mode traverses this region, contributing phase space values between 0.05 and 0.09 ps$^4$ rad$^{-4}$, though with a smaller impact on the total thermal conductivity. In the higher frequency region ($>$90 cm$^{-1}$), the phase space decreases progressively and contributes minimally to the total thermal conductivity, consistent with trends observed in anharmonic transport studies \cite{He2019}.  CuZn$_2$InTe$_4$ also follows the same behavior in the variation of phase space volume with frequency (see Fig.~\ref{fig2_a}b) as CuCd$_2$InTe$_4$, with a shift in the frequency regions. The scattering phase space is weaker, which is reflected in the lower scattering rate and higher thermal conductivity of CuZn$_2$InTe$_4$.  The above observations confirm that phonon softening and increased scattering phase space, particularly in the low-frequency region, are responsible for the suppression of lattice thermal conductivity.

\clearpage
\sloppy
\bibliography{references}
\end{document}